\documentclass[intlimits,twoside,a4paper]{article}

\usepackage{amsmath,amssymb}
\usepackage{graphicx}
\usepackage{wrapfig}

\usepackage[T2A]{fontenc}
\usepackage[cp1251]{inputenc}

\usepackage[eqsecnum]{cmpj2}

\issue{2012}{15}{2}{23002}

\doinumber{10.5488/CMP.15.23002}

\title[ Two-dimensional Yukawa fluids]%
{Phase equilibria and interfacial properties of two-dimensional Yukawa fluids}
\author[G.A.~M\'endez-Maldonado, M.~Gonz\'alez-Melchor, J.~Alejandre]{G.A.~M\'endez-Maldonado\refaddr{label1}, M.~Gonz\'alez-Melchor\refaddr{label2}, J.~Alejandre\refaddr{label3}\thanks{E-mail: jra@xanum.uam.mx}}

\addresses{
\addr{label1} Facultad de Ciencias F\'isico-Matem\'aticas,
Ben\'emerita Universidad Aut\'onoma de Puebla,\\
 Apartado Postal 1152, 72570, Puebla,  M\'exico
\addr{label2} Instituto de F\'{\i}sica ``Luis Rivera Terrazas'', Benem\'erita Universidad
Aut\'onoma de Puebla, \\Apartado Postal J--48, 72570, Puebla, M\'exico
\addr{label3} Departamento de Qu\'imica, Universidad Aut\'onoma
Metropolitana-Iztapalapa, \\Av. San Rafael Atlixco 186, Col. Vicentina,
09340, M\'exico Distrito Federal, M\'exico
}

\date{Received February 23, 2012}

\authorcopyright{G.A.~M\'endez-Maldonado, M.~Gonz\'alez-Melchor, J.~Alejandre, 2012}

\begin{document}

\maketitle


\begin{abstract}
Molecular dynamics simulations of two-dimensional soft Yukawa fluids are performed to analyze
the effect that the range of interaction has on coexisting densities and line tension. The attractive
 one-component fluid and equimolar mixtures containing positive and negative particles are studied at
different temperatures to locate the region where the vapor-solid and vapor-liquid phases are stable.
When the range of interaction  decreases, the critical temperature of the attractive one-component systems decreases.
However, for the charged mixtures it increases, and this opposite behaviour is understood in terms of the repulsive
interactions which are dominant for these systems.  The stable phase diagram of two-dimensional fluids is
 defined for smaller values of the decay parameter $\lambda$  than that of fluids in three dimensions.
The two-dimensional attractive one-component fluid has stable liquid-vapor phase diagram
for values of $\lambda<3$,  in contrast to the three-dimensional  case, where stability
has been observed even for values of $\lambda<15$. The same trend is observed
in equimolar mixtures of particles carrying opposite charges.


\pacs 05.70.Fh, 64.70.F-, 65.20.De, 68.03.Cd
\end{abstract}

\section{Introduction}

The location of stable phases in systems with molecular interactions has been a subject of great interest in many applications such as simple liquids, electrolyte solutions, colloidal suspension and water. It is well known that fluids with only repulsive interactions have a phase diagram that shows fluid-solid transition \cite{alder}. Attractive forces are needed to develop a vapor-liquid phase separation \cite{sw}. Several methods, among them, molecular simulations, theory of liquids and experiments are used to obtain information about the phase diagram of atomic and  molecular fluids. Although it is possible nowadays to obtain the phase boundaries using atomistic simulations of molecules including internal degrees of freedom  \cite{nerd,interfases}, from theoretical and simulation points of view, it is more convenient to use simple potential models such as square well (SW), Lennard-Jones (LJ), Yukawa and Coulomb  interactions. The effect that parameters of the potential has on the phase diagram in general, and in particular, on the critical properties can be systematically analyzed by computer simulations. The SW, LJ and attractive hard-core Yukawa (AHCY) potentials have a short-ranged repulsion and the attraction is long or variable-ranged. For these models it has been established in three-dimensional (3D) systems that the critical temperature decreases as the attraction decreases  \cite{vega92,smit92,lomba97,dijkstra02,orea07}. For very small ranges of attraction, when the vapor-liquid or liquid-solid phase diagrams for the AHCY model become metastable, it has been found that the critical point is below the triple point \cite{dijkstra02}.

The phase diagram and interfacial properties of ionic fluids, where the interaction has attraction and repulsion, has also been studied in 3D for the restricted primitive model (an electroneutral equimolar mixture of particles having the same size and carrying opposite charges)  \cite{rpm,rpm1} and for ionic fluids with asymmetry in size and charge \cite{depablo,pana,bresme}. The effect of range of interaction on the phase diagram of ionic fluids has been analyzed using the Yukawa potential. Mier-y-Ter\'an et al.  \cite{lmyt98} used the mean spherical approximation theory to calculate the critical properties of mixtures containing positive and negative particles at different range of interaction. They predicted that as the range of interaction decreases the critical temperature increases, it reaches a  maximum and then decreases. That finding is counterintuitive in the sense that less attractive potentials increase the critical temperature. The surface tension for mixtures of charged particles using the Yukawa model have also been reported using molecular dy\-na\-mics \cite{minerva04}. Fortini et al.  \cite{fortini06} performed computer simulations to locate the regions where the vapor-liquid and liquid-solid were stable. Their conclusions were that stable liquid-vapor phase diagrams were found for values of the screening parameter $\lambda  < 4$, in disagreement with the results of Caballero et al.  \cite{caballero06} where the phase diagram was stable for $\lambda < 10$. The results of Caballero et al. were obtained for  a slightly different potential, but according to Fortini et al., the difference in the results were not related with the potential. Hynninen and Panagiotopoulos  \cite{hynninen07} found that vapor-liquid phase transition of highly charged colloids is metastable with respect to the vapor-solid  phase diagram because at high temperatures the interaction becomes purely repulsive.

Apart from the work in 3D, it is also interesting to understand the phase behavior of ionic particles in two-dimensions. Experimental results on quasi two-dimensional (2D) colloidal suspensions show interesting properties  \cite{grier96,arauz98}. The phase diagrams of 2D systems obtained by computer simulations are scarce. The vapor-liquid and fluid-solid phase diagrams of the LJ fluid in two dimensions were obtained by Barker et al.  \cite{barker81} in 1981 using a liquid-state perturbation theory and Monte Carlo simulations. They found that the phase diagram was qualitatively similar to the 3D system. Santra et al.  \cite{santra08} in 2008 studied the nucleation rate of a liquid phase. They used the LJ model and Monte Carlo simulations to validate the predictions of the classical nucleation theory. They calculated for the first time the line tension of the 2D LJ fluid. Later on, Santra and  Bagchi  \cite{santra09} obtained the vapor-liquid phase diagram and line tension of the LJ model at different temperatures. The critical temperature from simulations for the 3D LJ is around 2.5 times greater than its value in 2D. The restricted primitive model of ions in two dimensions was studied by Weis et al.  \cite{weis98} using Monte Carlo simulations and integral equations theory. For this system the critical temperature in the 3D model is around 1.3 times its value in 2D. Analyzing the stability of phase diagrams in 2D might be useful to understand the phase separation in 3D and to investigate new phenomena which are not found in 3D.

The main goal of this work is to analyze the effect that the range of interaction has on coexisting densities and line tension of 2D fluids that interact with the soft Yukawa model. To our best knowledge, there have neither been  reported any phase diagrams nor surface tension for these fluids. The one-component systems with attractive interactions and the two-component mixture of positive and negative particles are studied at different ranges of interaction.

This work is organized as follows: The potential model and definition of the calculated properties  are presented in section~2. Results are discussed in section~3 and finally Concluding remarks and References are given.

\section{Potential  model and calculated properties}

The soft Yukawa potential is used to simulate the two-dimensional fluids of pure attractive spheres and equimolar mixtures of particles carrying the opposite charge,
\begin{equation}
\label{syuka}
U(r) = \left[\left( \frac{\sigma}{r}\right)^{225}+
\frac{q_{\alpha}q_{\beta} \re^{-\lambda(r/\sigma-1)}}{r/\sigma}\right]f_{\mathrm{min}}
\end{equation}
where  $\lambda^{-1}$ is a measure of the range of interaction
in dimensionless units and $q_{\alpha}$ is the charge of  particle $\alpha$ in the mixture case.
The particles are all the same size $\sigma$. For the attractive one-component
fluid $q_{\alpha}q_{\beta}=-1$. The  factor $f_{\mathrm{min}}=1$.075 is included to
have a potential which is zero at $r=\sigma$ and close to $-1$ at the minimum
for the attractive pairs. The short-ranged soft model, instead of the hard sphere model,
has been used earlier to study interfacial properties of the restricted primitive
model  \cite{rpm1,ions2} and ions with asymmetry in size and charge \cite{ions3}.
The results for the soft model were found to be in good agreement with those where
the short range repulsion is given by the hard sphere potential.
The advantage of the soft model is that it is straightforward to use molecular dynamics of continuous models.

Reduced units are used in this work for the two-dimensional systems: distance $r^*=r/\sigma$, energy $u^*=u/\epsilon$ (where $\epsilon  = U(r_{\mathrm{min}}))$, temperature $T^*=kT/\epsilon$,  density $\rho^*=\rho \sigma^2$, time $\Delta t^*= \Delta t (\epsilon/m\sigma^2)^{1/2}$, pressure $p^*=p\sigma^2/\epsilon$ and line tension $\gamma^*=\gamma \sigma/\epsilon$.
The density profile, $\rho(x)$, was obtained as,
\begin{equation}
\rho(x)=\frac{\langle N(x,x+\Delta x)\rangle }{\Delta A}
\end{equation}
where $\langle N(x,x+\Delta x)\rangle$ is the average number of particles with position between $x$ and $x+\Delta x$ and $\Delta A$ is the area of a slab.

The line tension of a planar  interface, using the mechanical definition of the atomic pressure,  \cite{lopez02} is
\begin{equation}
\gamma=0.5L_x\left[\left< P_{xx} \right >-\left< P_{yy} \right>\right],
\end{equation}
where  $L_x$ is the length of the simulation cell in the longest
direction and $P_{\alpha\alpha}$ $(\alpha=x,y)$ are diagonal components
 of the pressure tensor. The factor 0.5 outside the squared brackets takes into account the two symmetrical interfaces in the simulation.

The component $P_{xx}$ of the pressure tensor was calculated as,
\begin{equation}
P_{xx} A = \sum_i m_i v_{xi}^2 + \sum_i \sum_{j>i} {\bf F}_{ij} \cdot {\bf r}_{ij}\,,
\end{equation}
where $v_{xi}$ and $m_i$  are the velocity in the $x$ direction and the mass of particle $i$, respectively,  $A$ is the area of the system and ${\bf r}_{ij}={\bf r}_i - {\bf r}_j$  with ${\bf r}_i$ being the position of particle $i$. A similar expression for $P_{yy}$ was used. The force between particles $i$ and particle $j$ is,
\begin{equation}
{\bf F}_{ij}=-\frac{\partial u(r_{ij})}{\partial r_{ij}} \frac{{\bf r}_{ij}}{r_{ij}}\,.
\end{equation}

\section{Results}

Extensive molecular dynamics simulations, with a parallel program, were performed on particles interacting with the soft Yukawa potential at different ranges of interaction $\lambda^{-1}$. All the simulations to study inhomogeneous systems were carried out in non-squared simulation cells, keeping the total density and temperature (NAT)  constant.

\subsection{One-component systems}

Results for the one-component systems where particles interact with the attractive soft Yukawa potential were obtained for values of the reciprocal range of interaction $\lambda=1$, 1.8, 3, 4 and 6. The value of $\lambda=1.8$ was chosen to make a direct comparison with the 2D LJ fluid. It was shown in the previous work  \cite{yuka3d}  that in 3D the hard core Yukawa model with $\lambda=1.8$ gave equivalent results with the LJ potential.

Initially a number of 400 particles were placed in the middle of the simulation cell at a reduced density of 0.4 and the velocities randomly distributed. The dimensions of the simulation cell were $L_x=60\sigma$ and $L_y=15\sigma$. The equations of motion were solved using the velocity Verlet algorithm with a reduced time step of  $\Delta t^*=0.0005$. The temperature was kept constant with a global thermostat using a Nos\'e-Hoover chains of 4 thermostats  \cite{tuckerman} with parameter of 0.01.The cut-off distance was $6\sigma$ for all the simulations of pure fluids. The systems evolved to reach the equilibrium state where a liquid slab was surrounded by vapor \cite{ions2}. The systems in all cases were followed by 80 blocks of $10^6$  time steps.

\begin{figure}[!h]
\includegraphics[width=0.48\textwidth]{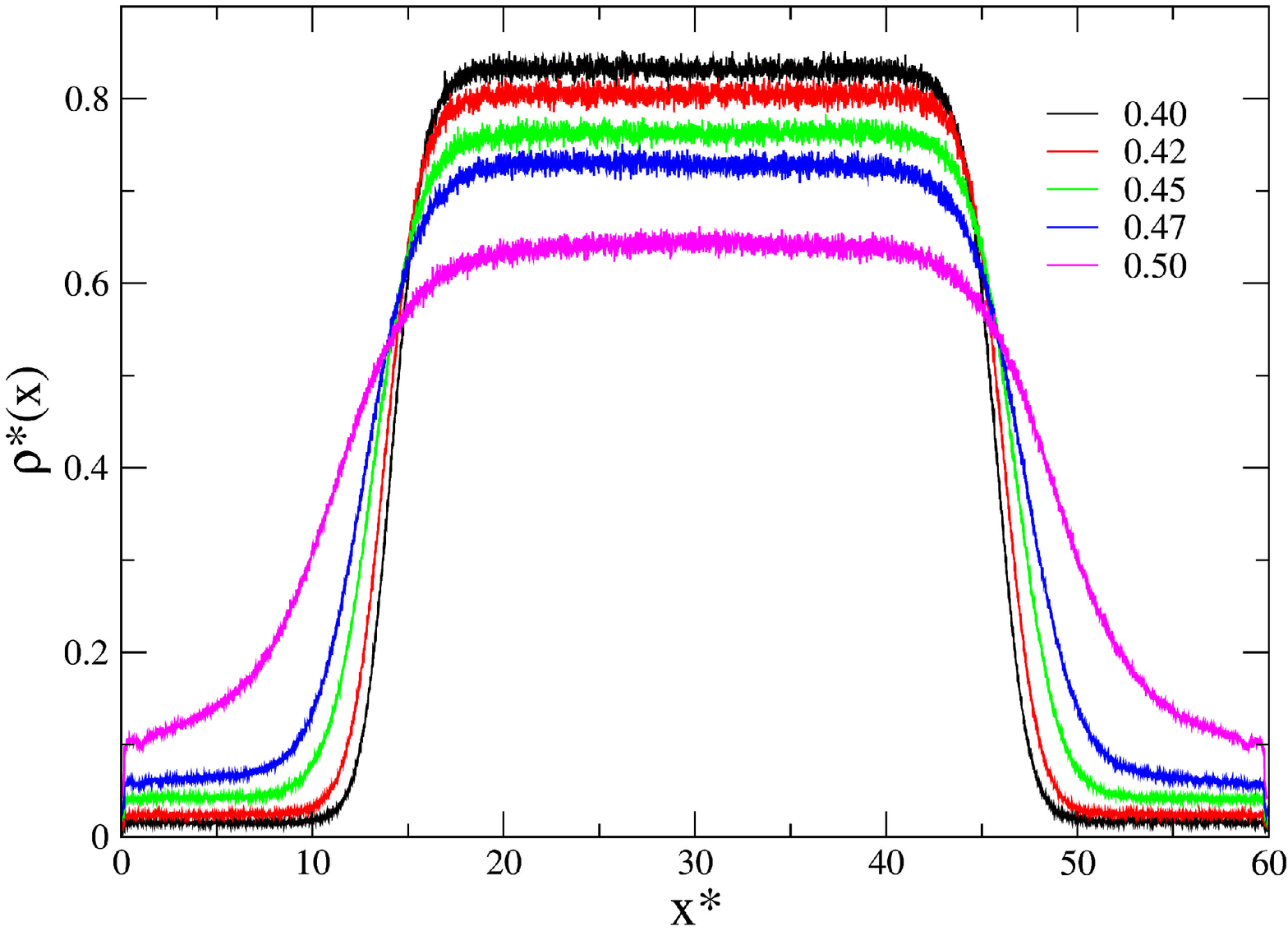}%
\hfill%
\includegraphics[width=0.48\textwidth]{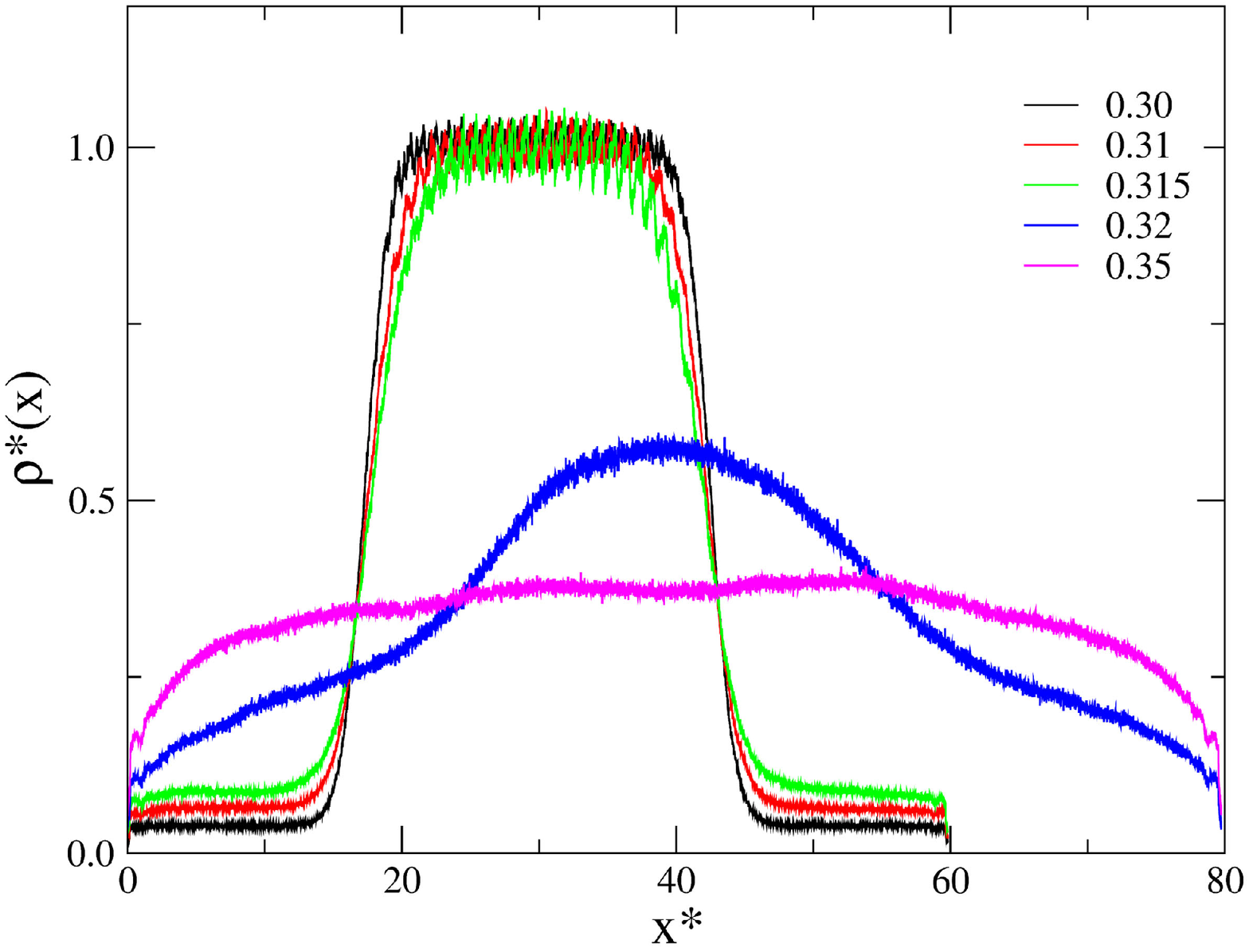}%
\\%
\parbox[t]{0.48\textwidth}{%
\caption{%
(Color online) Density profiles for attractive soft Yukawa one-component systems with $\lambda=1.8$ at different reduced temperatures. }
\label{perfiles1}%
}%
\hfill%
\parbox[t]{0.48\textwidth}{%
\caption{%
(Color online) Density profiles for attractive soft Yukawa one-component systems with $\lambda=4$ at different reduced temperatures. }
\label{perfiles2}%
}%
\end{figure}

The density profiles for $\lambda=1.8$ are shown in figure~\ref{perfiles1}. Two symmetrical interfaces are observed with a large amount of particles in the liquid and vapor regions which allows one to obtain the corresponding coexisting densities. Similar density profiles are found for $\lambda=1.0$, 3.0 and 6.0 not shown.

The density profiles for $\lambda=4$ are shown in figure~\ref{perfiles2}. There was not observed a vapor-liquid phase separation but vapor-solid equilibrium in a very narrow range of temperatures, from 0.30 to 0.315. Large density fluctuations were found at a reduced temperature of 0.32 but a well defined vapor-liquid interface was not stabilized. Homogeneous fluids were found for reduced temperatures above 0.35. The longest size of the simulation cell was increased in order to check if the phase separation from vapor-solid to vapor-liquid was not related to finite size effects. The two-dimensional attractive soft Yukawa model seems to have a stable vapor-liquid equilibrium for values of $\lambda$ around 3, this  is contrary to the 3D case where the same phase equilibrium is stable for values of $\lambda$ less than 15  \cite{orea10}.

\begin{figure}[!h]
\includegraphics[width=0.48\textwidth]{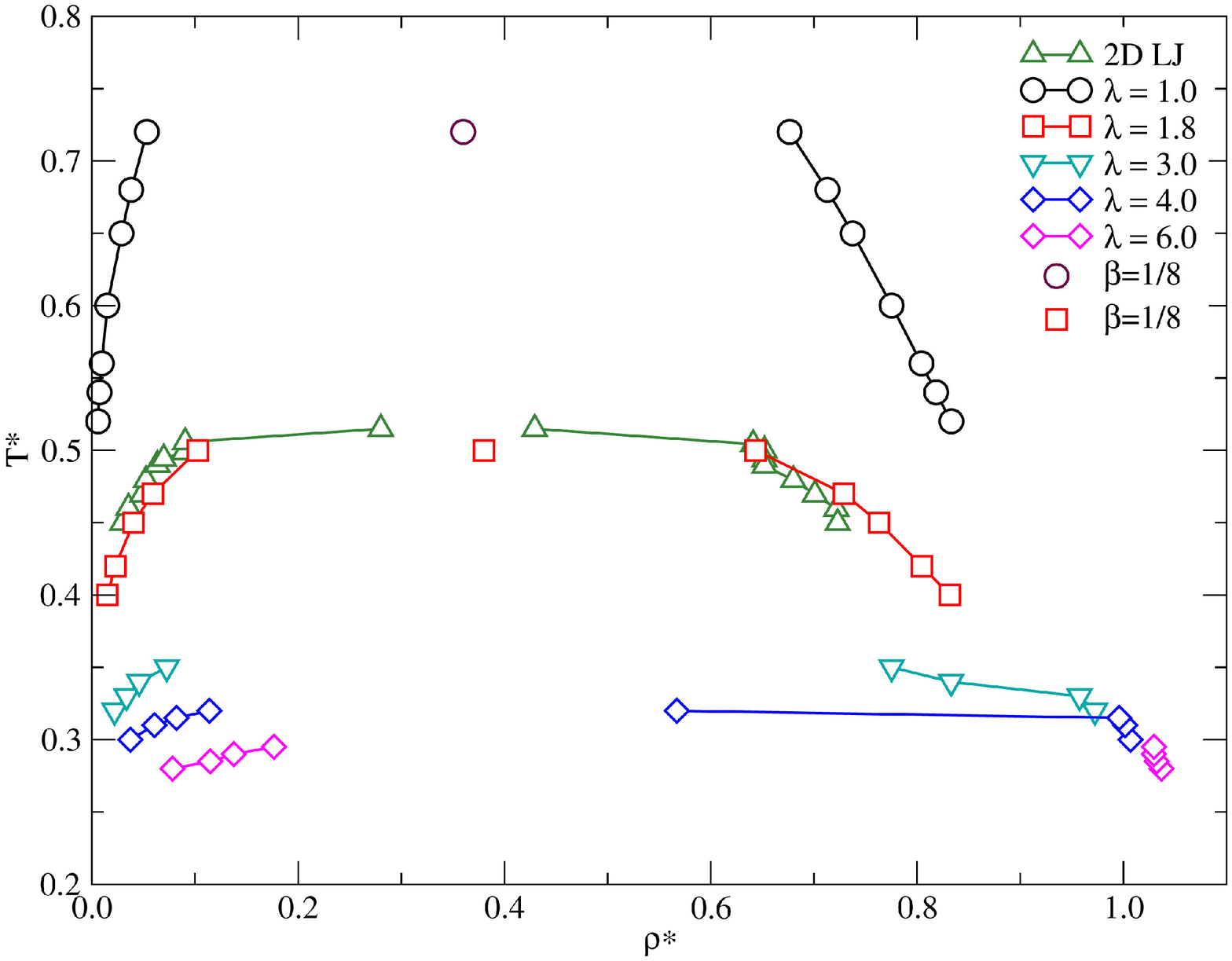}%
\hfill%
\includegraphics[width=0.48\textwidth]{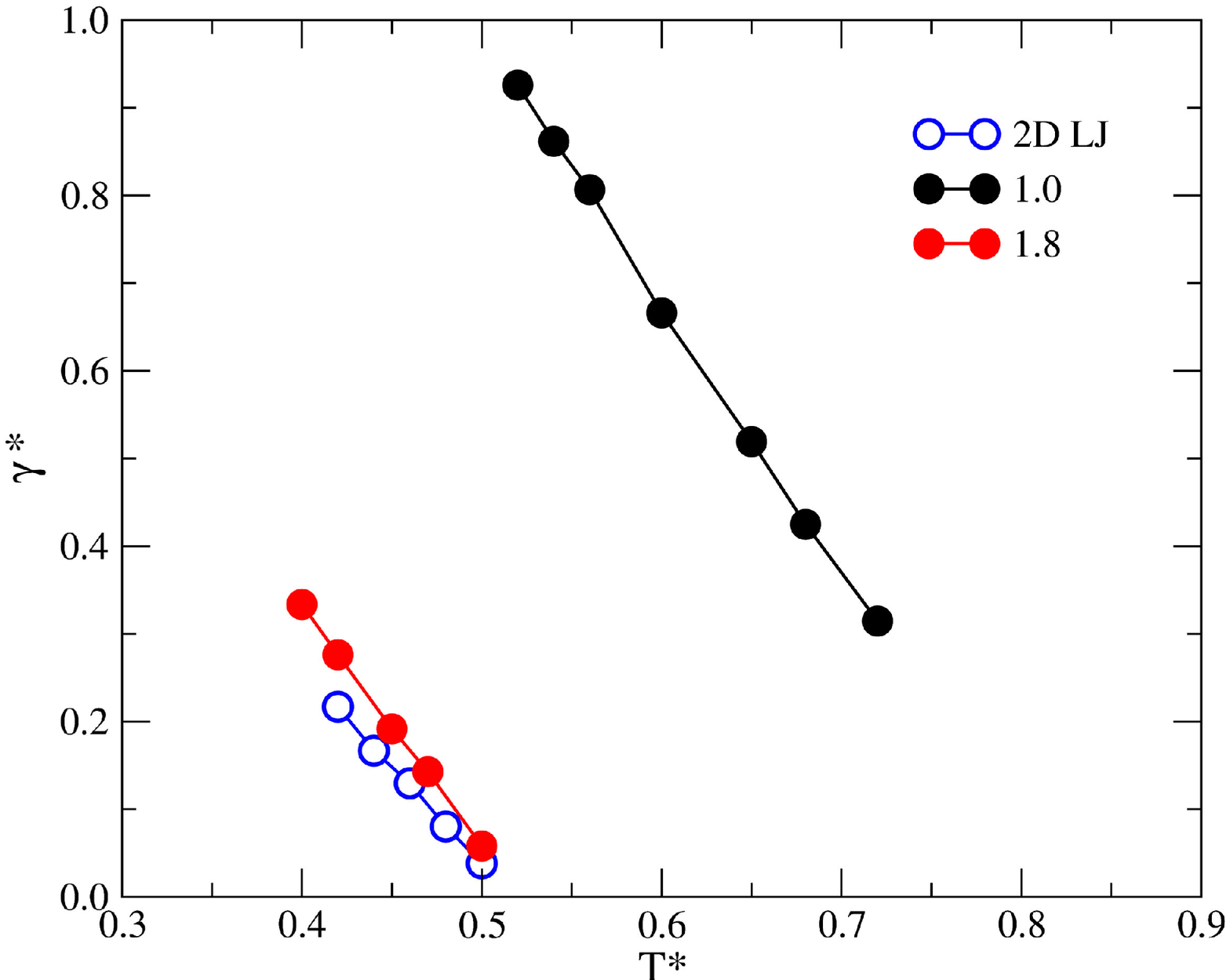}%
\\%
\parbox[t]{0.48\textwidth}{%
\caption{%
(Color online) Liquid-vapor phase diagram for attractive one-component systems with the inverse of the interaction range $\lambda=1.0$, 1.8, 3.0, 4.0 and 6.0. }
\label{lvyuka}%
}%
\hfill%
\parbox[t]{0.48\textwidth}{%
\caption{%
(Color online) Line tension for attractive one- component systems with $\lambda=1.0$ and 1.8 are shown with filled circles. The results for the 2D Lennard-Jones model  \cite{santra09} are shown with open circles.}
\label{styuka}%
}%
\end{figure}
The phase diagram of the attractive soft Yukawa potential is shown in figure~\ref{lvyuka} for different values of $\lambda$. As expected, the critical temperature decreases as the range of interaction decreases. At low temperatures the vapor-solid equilibrium is observed for values of $\lambda=1.8$ and 4. The vapor-liquid transition was not found for $\lambda=4$, it is metastable with respect to the vapor-solid equilibrium.
The critical density and temperature for $\lambda=1$ and $\lambda=1.8$ were estimated by fitting the coexisting densities to a rectilinear diameter law with a critical exponent of 1/8 according to a two-dimensional Ising model system  \cite{ising,ising2} and experimental results for methane  \cite{betaexp}. The final results were (0.36,0.72) and (0.38,0.50), respectively. The critical temperature using the coexisting densities for $\lambda=1.0$  might be around $T^*=0.8$ but the estimated value using the critical exponent of 1/8 is much smaller. The 2D soft Yukawa with $\lambda=1.8$ and LJ models give nearly the same results as shown in figure~\ref{lvyuka}.

The line tension of the attractive soft Yukawa model is shown in figure~\ref{styuka} as a function of temperature, its decay almost following a linear function. The critical temperature can be obtained when line tension is zero, the estimated values for $\lambda=1.0$ and $\lambda=1.8$ are 0.53  and 0.83, respectively.
The result for $\lambda=1$ is quite different from the value obtained using the coexisting densities and critical exponent of 1/8. The results for the two-dimensional LJ model, also shown in the figure, are in good agreement with the attractive soft Yukawa model with $\lambda=1.8$.

The phase diagram and line tension of 2D and 3D attractive soft Yukawa fluids are completely different. The vapor-liquid phase diagram in 3D is stable for values of $\lambda<15$, i.e., for very short ranges of interaction, whereas in the 2D case the stability is found for $\lambda<3$. Clearly, the strength of the global attraction needed in 2D to produce the separation is quite different from that in 3D.

\subsection{Mixture of particles carrying opposite charges}

Molecular dynamics simulations on equimolar binary mixtures of equal-size particles carrying opposite charges are carried out to analyze the effect that the range of interaction has on phase stability and line tension. The systems contained 1000 particles in a non-squared simulation cell of $L_x=150\sigma$ and $L_y=38\sigma$ dimensions. The potential was truncated at 15$\sigma$ and the reduced time step was 0.0005. The simulation protocol to obtain the coexisting densities and line tension was the same as that used for the attractive soft Yukawa model described above.  The systems were followed for at least 100 blocks of  $10^6$ time steps. The average properties were obtained from the last 40 blocks.

\begin{figure}[!h]
\centerline{\includegraphics[width=0.48\textwidth]{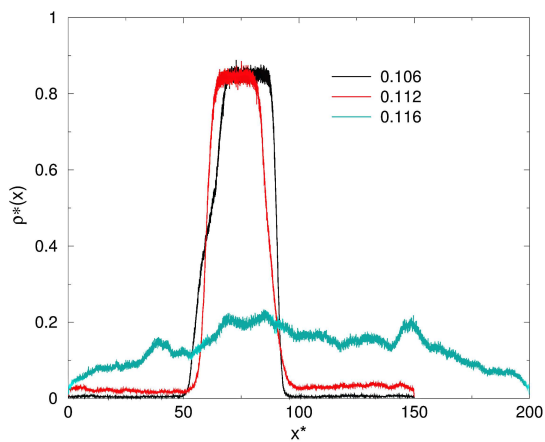}}
\caption{(Color online) Density profiles for the soft Yukawa mixture of charged particles with $\lambda=2.0$. The temperatures are given in the figure.}
\label{perfilesmix2}
\end{figure}
The phase diagram for the same system but in 3D has been calculated  \cite{fortini06} and it was found that  the vapor-liquid equilibrium was stable for values of $\lambda<4$.  In order to find the region where the vapor-liquid transition occurred, several simulations were performed in this work using $\lambda=2$. The difference in critical temperature for the restricted primitive model in 2D and 3D was less than 25 \% (see the Introduction). We expected to find the same trend in the binary mixtures using the soft Yukawa model. However, we did not find any evidence of vapor-liquid phase separation using the direct simulation of interfaces. A vapor-solid equilibrium was observed for reduced temperatures as low as 0.106. The amount of vapor increases when the reduced temperature rises to 0.112 but the solid phase  still remains stable. At $T^*=0.116$ the particles behave as a homogenous fluid, see figure~\ref{perfilesmix2}. As in the 3D mixture of charged particles with opposite sign interacting with the hard core Yukawa model, the critical temperature increases as the range of interaction decreases  \cite{lmyt98,fortini06}, i.e., the system increases its liquid density as $\lambda$ increases. In fact, for values of $\lambda>4$, the vapor-liquid phase diagram is meta-stable with respect to the vapor-solid equilibrium. In the same way, the possibility of finding a vapor-liquid phase transition in 2D charged mixtures would be in the direction of increasing the range of interaction.

\begin{figure}[ht]
\centerline{\includegraphics[width=9cm]{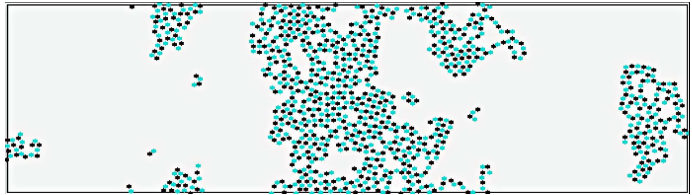}}
\caption{(Color online) Snapshot for the soft Yukawa mixture of charged particles with $\lambda=1.0$ and $T^*=0.079$. }
\label{foto09}
\end{figure}

Therefore, MD simulations were performed using $\lambda=1.0$. In this case, a  vapor-liquid phase separation was found in a very narrow range of reduced temperatures, from 0.078 to 0.08. The liquid contains alternated particles with opposite charges and large voids are observed, see figure~\ref{foto09} for $T^*=0.79$. In the vapor phase, the particles contain a large cluster and some particles form linear chains and rings.

\begin{figure}[ht]
\centerline{\includegraphics[width=0.48\textwidth]{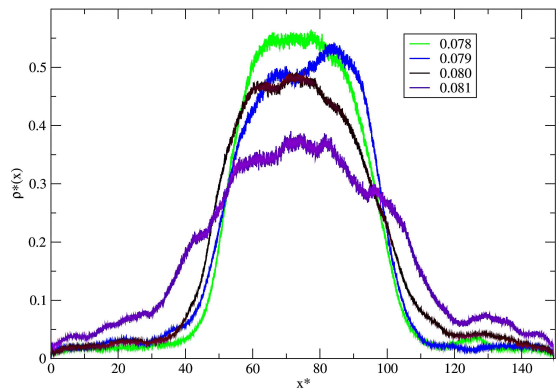}}
\caption{(Color online) Density profiles at different temperatures for the soft Yukawa mixture of charged particles with $\lambda=1.0$.}
\label{perfilesmix1}
\end{figure}

The density profiles are shown in figure~\ref{perfilesmix1} and have large fluctuations. However, the liquid and vapor can be estimated when simulations are run for several millions of configurations. The analysis of the radial distribution function for the region with higher density shows a behavior of a liquid.

\begin{figure}[ht]
\centerline{\includegraphics[width=0.48\textwidth]{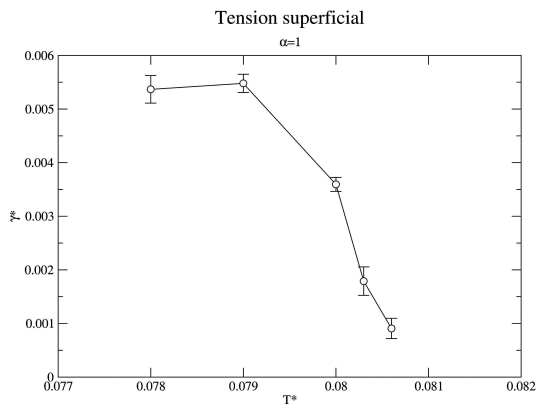}}
\caption{Line tension as a function of temperature for soft Yukawa mixture of charged particles with $\lambda=1.0$.}
\label{tension1}
\end{figure}

 The line tension for binary mixtures was also calculated and the results are shown in figure~\ref{tension1}. As found in simple fluids, the line tension decays with temperature.

\section{Concluding remarks}

The main conclusion found in this work is that the stability of vapor-liquid phase diagram in 2D fluids requires longer ranges of interactions than in 3D systems in both the one-component soft Yukawa model and the binary mixture of soft spheres carrying charges with the opposite sign. The range of interaction for these models increases when the inverse screening parameter $\lambda$ decreases. The metastable vapor-liquid transition, with respect to the vapor-solid equilibrium, for the one component model in 2D is found for values of $\lambda>4$ while in the 3D case it is found for $\lambda>15$, i.e., for very short ranges of interaction. For the binary mixture in 2D, metastability occurs for $\lambda>1$ in contrast to $\lambda>4$ observed in the 3D case. A possible explanation might be given using the classical nucleation theory \cite{cnt} arguments. The formation of a liquid phase in that theory requires that particles in a metastable vapor phase should  nucleate up to a critical size. The number of particles in a nucleus of 2D fluids for the same range of interaction is smaller than in 3D because in 3D the particles are in a sphere while in 2D they are in a circle. The particles in 2D have to interact longer distances than in 3D for the nucleus to reach the critical size.

On the other hand, the vapor-liquid phase diagram and line tension for the one-component attractive soft Yukawa model are in good agreement with those obtained using the LJ in 2D.

 \section*{Acknowledgements}

G.A.M.M. and M.G.M. acknowledge financial support from CONACyT (Project 129034), VIEP--BUAP (GOMM--EXC10--I), PROMEP/103.5/07/2594 and CA F\'isica
Computacional de la Materia Condensada. J.A. thanks Conacyt (Project 81667)  for financial support and to Laboratorio de Superc\'omputo for time allocation.

\ukrainianpart

\title{Фазова рівновага і міжфазні властивості двовимірних Юкава-плинів}

\author{Ґ.А.~Мендес-Мальдонадо\refaddr{label1}, М.~Ґонсалес-Мельчор\refaddr{label2}, Х.~Алехандре\refaddr{label3}}

\addresses{
\addr{label1} Факультет фізико-математичних наук, Автономний
університет Пуебла, 72570, Пуебла, Мексика
\addr{label2} Інститут
фізики, Автономний університет Пуебла, 72570, Пуебла, Мексика
\addr{label3} Хімічний факультет, Автономний університет
Метрополітана-Істапалапа, 09340, Федеральний округ Мехіко, Мексика
}

\makeukrtitle

\begin{abstract}
\tolerance=3000%
Для того, щоб проаналізувати, як впливає область взаємодії на
співіснуюючі густини і лінійний натяг, здійснено симуляції методом
молекулярної динаміки двовимірних м'яких Юкава-плинів.  Притягальний
однокомпонентний плин та еквімолярні суміші, що містять позитивні і
негативні частинки, досліджувалися при різних температурах таким
чином, щоб визначити область, в якій є стійкими фази пара-тверде тіло
і пара-рідина.  Зі зменшенням області взаємодії зменшується критична
температура притягальних однокомпонентних систем. Проте для
заряджених сумішей вона зростає, і ця відмінність у поведінці
пояснюється наявністю відштовхувальної взаємодії, яка домінує в цих
системах.  Діаграма стійкої фази двовимірних плинів є отримана для
менших значень параметра затухання $\lambda$, ніж у випадку тривимірних
плинів.  Двовимірний притягальний однокомпонентний плин має стійку
фазову діаграму рідина-пара для значень $\lambda<3$, що відрізняється
від тривимірного випадку, для якого стійкість спостерігається навіть
для значень $\lambda<15$. Така ж тенденція спостерігається в
еквімолярних сумішах протилежно заряджених частинок.

\keywords фазова діаграма, Юкава-плини, міжфазні властивості

\end{abstract}

\end{document}